\documentclass[letter]{ptptex}

\usepackage{graphicx}

\newcommand{\beq}{\begin{equation}}
\newcommand{\beqa}{\begin{eqnarray}}
\newcommand{\eeq}{\end{equation}}
\newcommand{\eeqa}{\end{eqnarray}}

\title{
A Null Test of the Cosmological Constant%
}

\author{
Takeshi \textsc{Chiba}$^{1}$ and Takashi \textsc{Nakamura}$^{2}$
}

\inst{
$^1$Department of Physics, College of Humanities and Sciences, \\
Nihon University, 
Tokyo 156-8550, Japan\\
$^2$Department of Physics, Kyoto University, 
Kyoto 606-8502, Japan
}

\recdate{July 25, 2007}

\abst{
We provide a consistency relation between cosmological observables 
in general relativity with the cosmological constant. 
Breaking of this relation at any redshift would imply the breakdown of 
the hypothesis of the cosmological constant as an explanation of 
the current acceleration of the universe. 
}

\begin{document}

\maketitle

\section{Introduction}

Recently,  one of us provided an explicit relation between the luminosity 
distance and the growth rate of density perturbations which should hold 
under the assumptions that general relativity is the correct theory 
of gravity and that the dark energy clustering is negligible\cite{ct}. 
In deriving the consistency 
relation, the Friedmann equation is not used, and hence 
the equation of state of dark energy, $w(z)$, is not assumed. 
$w(z)$ can be determined once the Friedmann equation (Einstein equation) 
is used \cite{nc,ht,alex}. 
Then, one may wonder whether another consistency relation, 
which is precisely zero for the cosmological constant, can be obtained 
if $w=-1$ is assumed, in such a way that the relation is precisely zero 
for the cosmological constant. The goal of this short note is to derive 
such a relation and with it to propose a null test for the cosmological constant. 
The basic idea is very simple: compare the matter density parameters determined 
from distance measurements assuming flat universe with the cosmological constant 
($\Omega_{M0}^{w-1}$ hereafter)  and 
those determined from other methods which are insensitive to the equation of state. 
We shall quantify the statement in the following. 

\section{Equation of State of Dark Energy from Observations}

First, from distance measurements, in terms of the coordinate distance $r(z)=d_L(z)/(1+z)$,  
the Hubble parameter is rewritten in a flat universe as \cite{nc,ht,alex}
\beqa
H(z)^2=\frac{1}{r'(z)^2},
\label{hubble:dl}
\eeqa
where the prime denotes the derivative with respect to $z$. 
Then, using the Friedmann equation for a flat universe,
\beqa
H(z)^2=H_0^2\Omega_{M0}(1+z)^3+\frac{8\pi G}{3}\rho_{\rm x}(z),
\eeqa
we obtain
\beqa
8\pi G\rho_{\rm x}(z)=\frac{3}{r'(z)^2}
-3H_0^2\Omega_{M0}(1+z)^3,
\label{friedmann}
\eeqa
where $\Omega_{M0}$ is the present matter density parameter and $\rho_{\rm x}(z)$ 
is the energy density  of dark energy. 
{}From the time derivative of Eq. (\ref{friedmann}) and the energy-momentum 
conservation of dark energy, $\dot\rho_{\rm x}+3H(1+w)\rho_{\rm x}=0$, we obtain
\beqa
(1+w(z))8\pi G\rho_{\rm x}(z)=
-2\frac{(1+z)r''(z)}{r'(z)^3}-3H_0^2\Omega_{M0}(1+z)^3.
\label{pressure}
\eeqa
Using Eqs.(\ref{friedmann}) and (\ref{pressure}),  
$w(z)$ may be written as \cite{nc,ht,alex}
\beqa
1+w(z)=\frac{2(1+z)r''(z)+3H_0^2\Omega_{M0}r'(z)^3(1+z)^3}
{3r'(z)\left(H_0^2\Omega_{M0}r'(z)^2(1+z)^3-1\right)}.
\label{w}
\eeqa
Note that we do not assume any functional form of $w(z)$. 
At this stage, $w(z)$ determined from distance measurements exhibits 
degeneracy with  $\Omega_{M0}$. However, $\Omega_{M0}h^2$  can be determined from 
measurements of CMB anisotropies \cite{wmap} 
being insensitive to the equation of state.  
Moreover,  from the evolution equation of $\delta(z)$ derived from the fluid equations,
\beqa
\ddot \delta+2H\dot\delta-\frac{3}{2}H_0^2\Omega_{M0}(1+z)^3\delta=0,
\label{linear:gr}
\eeqa 
$\Omega_{M0}$  can be written in terms of a 
density perturbation $\delta(z)$ 
independently of the equation of state 
(if dark energy is almost smooth) as \cite{alex} 
\beqa
\Omega_{M0}=\frac{\delta'(0)^2}{3}\left(\int^{\infty}_0\frac{\delta(z)}{1+z}
(-\delta'(z))dz\right)^{-1}.
\label{omega:delta}
\eeqa

\section{A Null Test of the Cosmological Constant}

{}From Eq. (\ref{w}), we define a function
\beq
\Omega_{M0}^{w=-1}(z)=-\frac{2r''(z)}{3 H_0^2(1+z)^2r'(z)^3} ,
\label{omega:dl}
\eeq
which coincides with $\Omega_{M0}$ for the cosmological constant. 
Equating  Eq. (\ref{omega:dl}) with Eq. (\ref{omega:delta}) thus gives a consistency 
relation for the cosmological constant, 
\beqa
\frac{\Omega_{M0}^{w=-1}(z)}{\Omega_{M0}}-1
&=&-\frac{2r''(z)}{3 \Omega_{M0}H_0^2(1+z)^2r'(z)^3} -1 \nonumber\\
&=&-\frac{2r''(z)}{ H_0^2(1+z)^2r'(z)^3\delta'(0)^2}\int^{\infty}_0\frac{\delta(z)}{1+z}
(-\delta'(z))dz-1=0 ,
\label{nulltest}
\eeqa 
which is the main result of this paper 
\footnote{The time derivative of Eq. (\ref{omega:dl}) for $w=-1$ gives the well-known 
relation of the jerk \cite{cn} for the flat universe with/without the cosmological constant: 
$\frac{a^2(d^3a/dt^3)}{(da/dt)^3}=1$. In this sense, 
Eq. (\ref{nulltest}) may be regarded as an integral form of the jerk relation. }
: compare $\Omega_{M0}^{w=-1}$ determined 
assuming a flat universe with the cosmological constant and $\Omega_{M0}$ 
 determined from other methods which are insensitive to the equation of state. 
$r(z)$ can be determined from distant measurements of type Ia supernovae (SNIa) \cite{snia}, 
gamma ray bursts (GRB) \cite{grb} and the baryon acoustic oscillation (BAO) 
in the matter power spectrum \cite{bao} and $\delta(z)$ can be determined from measurements 
of the weak lensing of galaxies \cite{kaiser} and of the evolution of cluster number density 
\cite{bahcall}, for example.   
If observational data indicate that the left-hand-side of Eq. (\ref{nulltest}) 
is nonzero at any redshift, 
this would be a clear signature of dynamical dark energy (or a non-flat universe). 
In this sense, Eq. (\ref{nulltest}) provides a null test of the cosmological constant.  

\begin{figure}
\includegraphics[width=13cm]{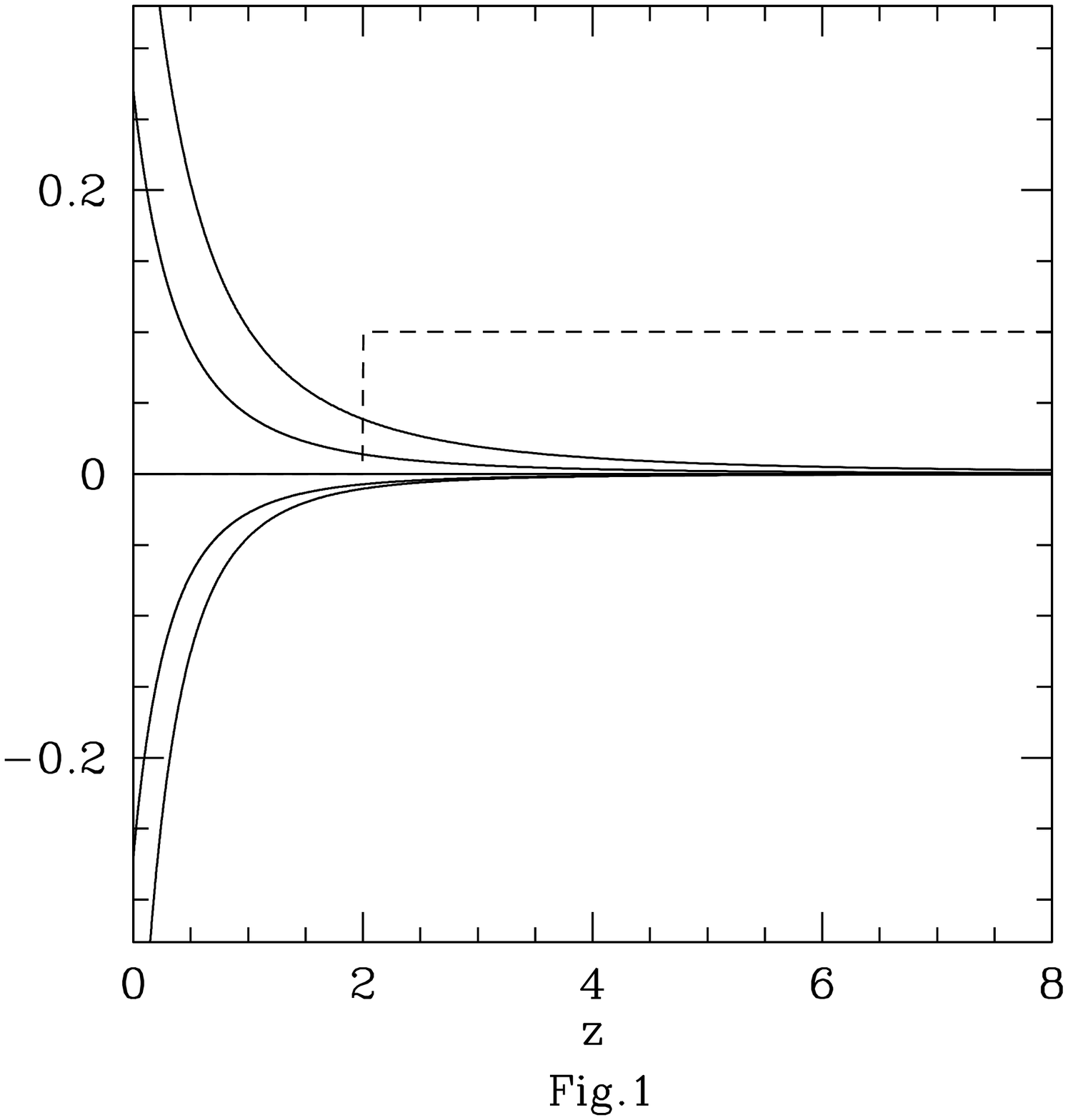} 
\caption{Left-hand side of Eq.(\ref{nulltest}) as a function of $z$ 
for several equation of state of 
dark energy. Solid lines are for $w=-0.8,-0.9,-1,-1.1,-1.2$ 
(from top to bottom) and the dashed line is for $w(z)=-1(z\leq 2),0(z>2)$. }
\label{fig1}
\end{figure}

In Fig. \ref{fig1}, the left-hand side of Eq. (\ref{nulltest}) is plotted for 
several values of $w(z)$: $w=-0.8,-0.9,-1,-1.1,-1.2$; $w(z)=-1(0)$ for $z\leq 2(z>2)$
\footnote{One may think that the constancy of $\Omega_{M0}^{w=-1}(z)$ determined from 
distance measurements alone for some redshift ranges could 
be a test of the cosmological constant. However, 
this second example shows explicitly that verifying $\Omega_{M0}^{w=-1}=\Omega_{M0}$ 
is crucial in testing the cosmological constant.}. 
Given $w(z)$, we compute $r(z)$ from Eq. (\ref{hubble:dl}) and $\delta$ from the evolution 
equation of $\delta$. A flat universe with $\Omega_{M0}=0.27$ is assumed. 
We note that Eq. (\ref{nulltest}) can be written in terms of 
$w(z)$ as
\beqa
\frac{\Omega_{M0}^{w=-1}(z)}{\Omega_{M0}}-1=\frac{1-\Omega_{M0}}{\Omega_{M0}}
(1+w(z))\exp\left(3\int^z_0\frac{w(z')}{1+z'}dz'\right).
\eeqa
If dark energy is not the cosmological constant, 
a deviation of more than 10\% is expected at lower ($z<1$) redshifts for 
a constant $w$. Moreover, even if dark energy behaves like the cosmological 
constant ($w=-1$) at lower redshifts, if it tracks matter ($w=0$) 
at higher redshifts, which is the case in certain quintessence 
models with exponential potentials \cite{exp}, then a 10\% deviation would be 
expected even at higher redshifts. Therefore, distance measurements at higher redshifts 
(by GRB and BAO) 
would be complementary to those at lower redshifts (by SNIa) in testing the cosmological 
constant. 

Currently from the measurements of SNIa, assuming a flat universe with $\Lambda$, 
$\Omega_{M0}^{w=-1}$ is determined to be $\Omega_{M0}^{w=-1}=0.29\pm^{0.05}_{0.03}$ 
\cite{riess04}.\footnote{The same analysis of the gold dataset of Riess et al.(2007)\cite{snia} 
gives a slightly higher $\Omega_{M0}^{w=-1}=0.34\pm^{0.05}_{0.03}$. Consequently, 
$\frac{\Omega_{M0}^{w=-1}}{\Omega_{M0}}-1=0.42\pm^{0.50}_{0.42}$. }
On the other hand, the measurements of CMB anisotropies by WMAP yields  
$\Omega_{M0}h^2=0.127\pm 0.008$ \cite{wmap}, which when combined with the Hubble parameter 
determined by the HST \cite{hst}, $h=0.72\pm0.08$,  gives $\Omega_{M0}=0.24\pm 0.05$. Hence 
$\frac{\Omega_{M0}^{w=-1}}{\Omega_{M0}}-1=0.21\pm^{0.46}_{0.38}$. 
The two $\Omega_{M0}$ coincides each 
other and the cosmological constant passes the null test.  
In any case, precision measurements of $\Omega_{M0}$ themselves 
can be a test of the cosmological constant.

\section*{Acknowledgements}
This work was supported in part by  Grants-in-Aid for Scientific 
Research [Nos.17204018 (TC) and 19035006 (TN)] from the Japan Society for the 
Promotion of Science and in part by Nihon University
and  by a Grant-in-Aid for the 21st Century COE
``Center for Diversity and Universality in Physics''
from the Ministry of Education, Culture, Sports, Science and Technology
(MEXT) of Japan. The numerical calculations were performed at YITP in Kyoto University.

%

\end{document}